\documentclass[11pt]{article}
\usepackage{latexsym}
\usepackage{graphicx}

\def\AFOUR{%
\setlength{\textheight}{8.5in}%
\setlength{\textwidth}{5.75in}%
\setlength{\topmargin}{-0.375in}%
\hoffset=-.5in%
\renewcommand{\baselinestretch}{1.17}%
\setlength{\parskip}{6pt plus 2pt}%
}

\AFOUR                                           

\expandafter\ifx\csname amssym.def\endcsname\relax \else\endinput\fi
\expandafter\edef\csname amssym.def\endcsname{%
       \catcode`\noexpand\@=\the\catcode`\@\space}
\catcode`\@=11
\def\undefine#1{\let#1\undefined}
\def\newsymbol#1#2#3#4#5{\let\next@\relax
 \ifnum#2=\@ne\let\next@\msafam@\else
 \ifnum#2=\tw@\let\next@\msbfam@\fi\fi
 \mathchardef#1="#3\next@#4#5}
\def\mathhexbox@#1#2#3{\relax
 \ifmmode\mathpalette{}{\m@th\mathchar"#1#2#3}%
 \else\leavevmode\hbox{$\m@th\mathchar"#1#2#3$}\fi}
\def\hexnumber@#1{\ifcase#1 0\or 1\or 2\or 3\or 4\or 5\or 6\or 7\or 8\or
 9\or A\or B\or C\or D\or E\or F\fi}

\font\tenmsa=msam10
\font\sevenmsa=msam7
\font\fivemsa=msam5
\newfam\msafam
\textfont\msafam=\tenmsa
\scriptfont\msafam=\sevenmsa
\scriptscriptfont\msafam=\fivemsa
\edef\msafam@{\hexnumber@\msafam}
\mathchardef\dabar@"0\msafam@39
\def\dashrightarrow{\mathrel{\dabar@\dabar@\mathchar"0\msafam@4B}}
\def\dashleftarrow{\mathrel{\mathchar"0\msafam@4C\dabar@\dabar@}}

\def\ulcorner{\delimiter"4\msafam@70\msafam@70 }
\def\urcorner{\delimiter"5\msafam@71\msafam@71 }
\def\llcorner{\delimiter"4\msafam@78\msafam@78 }
\def\lrcorner{\delimiter"5\msafam@79\msafam@79 }
\def\yen{{\mathhexbox@\msafam@55}}
\def\checkmark{{\mathhexbox@\msafam@58}}
\def\circledR{{\mathhexbox@\msafam@72}}
\def\maltese{{\mathhexbox@\msafam@7A}}
\def\circledS{{\mathhexbox@\msafam@73}}

\font\tenmsb=msbm10
\font\sevenmsb=msbm7
\font\fivemsb=msbm5
\newfam\msbfam
\textfont\msbfam=\tenmsb
\scriptfont\msbfam=\sevenmsb
\scriptscriptfont\msbfam=\fivemsb
\edef\msbfam@{\hexnumber@\msbfam}
\def\Bbb#1{{\fam\msbfam\relax#1}}
\def\widehat#1{\setbox\z@\hbox{$\m@th#1$}%
 \ifdim\wd\z@>\tw@ em\mathaccent"0\msbfam@5B{#1}%
 \else\mathaccent"0362{#1}\fi}
\def\widetilde#1{\setbox\z@\hbox{$\m@th#1$}%
 \ifdim\wd\z@>\tw@ em\mathaccent"0\msbfam@5D{#1}%
 \else\mathaccent"0365{#1}\fi}
\font\teneufm=eufm10
\font\seveneufm=eufm7
\font\fiveeufm=eufm5
\newfam\eufmfam
\textfont\eufmfam=\teneufm
\scriptfont\eufmfam=\seveneufm
\scriptscriptfont\eufmfam=\fiveeufm
\def\frak#1{{\fam\eufmfam\relax#1}}

\csname amssym.def\endcsname

\makeatletter
\def\section{\@startsection {section}{1}{\z@}{-3.5ex plus -1ex minus
 -.2ex}{2.3ex plus .2ex}{\large\sc}}
\def\subsection{\@startsection{subsection}{2}{\z@}{-3.25ex plus -1ex minus
 -.2ex}{1.5ex plus .2ex}{\normalsize\sc}}
\makeatother

\makeatletter
\@addtoreset{equation}{section}

\makeatother

\newcommand{\nc}{\newcommand}
\newcommand{\rnc}{\renewcommand}

\nc{\subs}[1]{{\vspace*{0.5cm}}%
{\noindent\underline{\small\sc #1}}{\addcontentsline{toc}{subsubsection}{#1}}%
{\vspace*{0.3cm}}}

\nc{\subss}[1]{{\vspace*{0.5cm}}%
{\noindent\underline{\small\sc #1}}%
{\vspace*{0.3cm}}}

\nc{\chap}[1]{{\clearpage}%
\begin{center}%
{\noindent\underline{\large\sc #1}}{\addcontentsline{toc}{section}{#1}}%
\end{center}%
{\vspace*{0.3cm}}}

\nc{\be}{\begin{equation}}
\nc{\ee}{\end{equation}}
\nc{\bea}{\begin{eqnarray}}
\nc{\eea}{\end{eqnarray}}

\nc{\trac}[2]{{\textstyle\frac{#1}{#2}}}

\nc{\ex}[1]{\mbox{e}^{\,\textstyle#1}}

\nc{\CC}{\Bbb{C}}
\nc{\HH}{\Bbb{H}}
\nc{\PP}{\Bbb{P}}
\nc{\RR}{\Bbb{R}}
\nc{\ZZ}{\Bbb{Z}}
\nc{\II}{\Bbb{I}}
\nc{\EE}{\Bbb{E}}

\rnc{\a}{\alpha}
\rnc{\b}{\beta}
\rnc{\d}{\delta}
\nc{\ga}{\gamma}
\nc{\la}{\lambda}
\nc{\f}{\phi}
\nc{\p}{\psi}
\nc{\e}{\eta}
\rnc{\c}{\chi}
\nc{\eps}{\epsilon}
\nc{\om}{\omega}
\nc{\Om}{\Omega}

\nc{\symx}{\circledS}
\newsymbol\smallsmile 1360
\newsymbol\smallfrown 1361
\nc{\ad}{\mathop{\mbox{ad}}\nolimits}
\nc{\tr}{\mathop{\mbox{tr}}\nolimits}
\nc{\Tr}{\mathop{\mbox{Tr}}\nolimits}
\nc{\Det}{\mathop{\mbox{Det}}\nolimits}
\rnc{\det}{\mathop{\mbox{det}}\nolimits}
\nc{\rk}{\mathop{\mbox{rk}}\nolimits}
\nc{\del}{\partial}
\nc{\diag}{\mathop{\mbox{diag}}\nolimits}
\nc{\ra}{\rightarrow}
\nc{\Ra}{\Rightarrow}
\nc{\LRa}{\Leftrightarrow}
\nc{\lra}{\leftrightarrow}
\nc{\ot}{\otimes}
\rnc{\ss}{\subset}
\nc{\nul}{\noindent\underline}
\nc{\non}{\nonumber\\}
\nc{\mat}[4]{\left(\begin{array}{cc}#1&#2\\#3&#4\end{array}\right)}
\rnc{\lg}{\frak{g}}
\nc{\G}[3]{\Gamma^{#1}_{\;{#2}{#3}}}
\nc{\nam}{\nabla_{\mu}}
\nc{\nan}{\nabla_{\nu}}
\nc{\dx}{\dot{x}}
\nc{\dxl}{\dot{x}^{\la}}
\nc{\dxm}{\dot{x}^{\mu}}
\nc{\dxn}{\dot{x}^{\nu}}
\nc{\ddx}{\ddot{x}}
\nc{\ddxm}{\ddot{x}^{\mu}}
\nc{\ddxn}{\ddot{x}^{\nu}}
\nc{\dxi}{\dot{\xi}}
\nc{\ddxi}{\ddot{\xi}}

\def\const{{\rm const}}

\def\la{\label}
\def \p {\phi}

\begin{document}

\rightline{SISSA/23/2004/EP}

\vfill

\begin{center}
{\Large\sc The Universality of Penrose Limits\\[.4cm]
near Space-Time Singularities}
\end{center}

\begin{center}
{\large
M.\ Blau${}^{a,}$\footnote{e-mail: {\tt matthias.blau@unine.ch}},
M. Borunda${}^{b,c,}$\footnote{e-mail: {\tt mborunda@he.sissa.it}},
M.\ O'Loughlin${}^{b,}$\footnote{e-mail: {\tt loughlin@sissa.it}},
G.\ Papadopoulos${}^{d,}$\footnote{e-mail: {\tt gpapas@mth.kcl.ac.uk}}
}
\end{center}

\vskip 0.05 cm
\centerline{\it ${}^a$ Institut de Physique, Universit\'e de Neuch\^atel,
Rue Breguet 1}
\centerline{\it CH-2000 Neuch\^atel, Switzerland}

\vskip 0.05 cm
\centerline{\it ${}^b$ ISAS -- SISSA, Via Beirut 2-4, I-34013 Trieste, Italy}

\vskip 0.05 cm
\centerline{\it ${}^c$ INFN, Sezione di Trieste, Italy}

\vskip 0.05 cm
\centerline{\it ${}^d$ Department of Mathematics,  King's College London}
 \centerline{\it London WC2R 2LS, U.K. }

\vskip -2.0 cm

\begin{center}
{\bf Abstract}
\end{center}

We prove that Penrose limits of metrics with arbitrary singularities
of power-law type show a universal leading $u^{-2}$-behaviour near the
singularity provided that the dominant energy condition is satisfied
and not saturated. For generic power-law singularities of this type
the oscillator frequencies of the resulting homogeneous singular plane
wave turn out to lie in a range which is known to allow for an analytic
extension of string modes through the singularity. The discussion is
phrased in terms of the recently obtained covariant characterisation of
the Penrose limit; the relation with null geodesic deviation is explained
in detail.

\vfill

\newpage
\begin{small}
\tableofcontents
\end{small}

\setcounter{footnote}{0}

\section{Introduction}

The results of \cite{bfhp1,rrm,bfhp2,bmn,mt} have led to renewed
interest in the Penrose limit construction \cite{penrose,gueven,bfp}.
The Penrose limit associates to every space-time metric $g_{\mu\nu}$ and
choice of null geodesic $\gamma$ in that space-time a plane wave metric
\be
ds^2=g_{\mu\nu}dx^{\mu}dx^{\nu} \ra 
2dudv + A_{ab}(u)x^a x^b du^2 + d\vec{x}^2\;\;.
\ee
Here $A_{ab}(u)$ is the plane wave profile matrix and the computation
of the Penrose limit along $\gamma(u)$ amounts to determining the matrix
$A_{ab}(u)$ from the original metric $g_{\mu\nu}$.

Recently, in \cite{bbop1} a simple covariant characterisation and
definition of the Penrose limit wave profile matrix $A_{ab}(u)$ was
obtained which does not require taking any limit and which shows that
$A_{ab}(u)$ directly encodes diffeomorphism invariant information about
the original space-time metric.  The geometric significance of $A_{ab}(u)$
(and hence of the Penrose limit) turns out to be that it is the standard
\cite[Section 4.2]{HE} {\em transverse null geodesic deviation matrix}
of the original metric along the null geodesic $\gamma(u)$.

The relevance of this result lies in the fact that it tells us precisely
which aspects of the original background, namely covariant information
about the rate of growth of curvature and geodesic deviation along a
null geodesic, are detected by the Penrose limit and hence probed by,
say, string theory in the resulting plane wave background.

In particular, as a first step towards studying string propagation in
singular (and perhaps time-dependent) space-time backgrounds, it is then
of interest to determine the Penrose limits of space-time singularities
in general. And here we find a pleasant surprise, namely a remarkably
universal behaviour of Penrose limits of space-time singularities.

It had already been found for a variety of particular brane and
cosmological backgrounds (see e.g.\ \cite{bfp,fis,patricot,kunze})
that the exact Penrose limit is characterised by a wave profile
of the special form
\be
A_{ab}(u) \sim u^{-2}\;\;.
\ee
Plane wave metrics with precisely such a profile have the scale
invariance $(u,v)\ra(\lambda u,\lambda^{-1}v)$ and are thus {\em
homogeneous} singular plane waves (HPWs) \cite{bfp,prt,hpw}. Without loss
of generality, $A_{ab}(u)$ can be chosen to be diagonal and, anticipating
the interpretation of the entries of $A_{ab}$ as harmonic oscilator
frequencies, we will parametrise $A_{ab}(u)$ as
\be
A_{ab}(u) = -\omega_a^2\d_{ab} u^{-2} \;\;,
\label{2}
\ee
where $\omega_a^2$ can be positive or negative.

Moreover, in \cite{mbictplec,bbop1} we observed that the 
Penrose limit of space-time singularities of cosmological FRW and
Schwarzschild-like metrics, i.e.\ the leading behaviour of the profile 
$A_{ab}(u)$ as one approaches the singularity,
is also of the above form. This led us to the
\begin{quote}
{\bf Conjecture:} Penrose limits of (in some suitable sense physically
reasonable)
space-time singularities are singular homogeneous plane
waves with wave profile $A_{ab}(u) \sim u^{-2}$.
\end{quote}
Given the above relation between Penrose limits and geodesic deviation,
this amounts to the conjecture that null geodesic deviation shows a
universal $u^{-2}$-behaviour near such space-time singularities.

The main result of this paper is the proof of this conjecture for a
large class of spherically symmetric space-time  singularities known as
``singularities of power-law type'' or Szekeres-Iyer metrics \cite{SI,CS}.
These can be spacelike, timelike or null singularities, and encompass
practically all known spherically symmetric singular solutions of the
Einstein equations.  For technical reasons we focus on the spacelike
and timelike singularities.

It is evident that, in order to be able to say anything of substance about
the behaviour near a singularity, some supplementary energy condition
has to be imposed. This is what we will do and, specifically, we will prove
that
\begin{quote}
Penrose Limits of spherically symmetric spacelike or timelike
singularities of power-law type satisfying (but not saturating) the
Dominant Energy Condition (DEC) are singular homogeneous plane waves of
the type (\ref{2}).
\end{quote}
Particle and wave propagation in the background (\ref{2}) exhibit a
qualitatively different behaviour for the `frequency squares' $\omega_a^2$
bounded by $1/4$ from above or below.  In all the explicit examples that
had been worked out one finds that $\omega_a^2$ 
is in the range $\omega_a^2 \leq 1/4$. 
We will show that this is indeed also the generic behaviour:
\begin{quote}
The resulting frequency squares $\omega_a^2$ are bounded from above by
1/4 unless one is on the border to an extremal equation of state.
\end{quote}
Here by ``extremal'' we mean, following the terminology of \cite{SI},
near-singularity energy-momentum tensors saturating the DEC, and
``border'' refers to a border in the Szekeres-Iyer phase diagrams
\cite{SI}, or Figure 2 in section 4.4 of the present paper.

The explicit proof and examples of metrics displaying a different, more
singular, behaviour illustrate that this universal $u^{-2}$-behaviour
is not simply a consequence of dimensional analysis. Rather, it is the
strong form of the DEC which guarantees that the singularity is no worse
than this and, in fact, precisely sufficiently benign (due to the bound on
the frequencies) to allow for a consistent string propagation through the
singularity (see e.g.\ \cite{sanchez,prt}). The fact that large classes of
physically reasonable metrics with space-time singularities give rise to
such a behaviour is certainly encouraging and perhaps somewhat unexpected.

In section 2 we discuss the geodesic deviation approach to Penrose
Limits. In section 2.1 we describe how to define and calculate the
transverse null geodesic deviation matrix. We also establish the
equivalence of the characterisation of $A_{ab}(u)$ in terms of the
Riemann tensor of the original metric, obtained in \cite{bbop1}, and the
description in terms of geodesic deviation we will use here. Section 2.2
contains some related comments on null congruences and solutions to
the Hamilton-Jacobi equations.  In section 2.3, as an illustration of
the geodesic deviation method, and as preparation for the calculations
of sections 3 and 4, we determine all the Penrose limits of a static
spherically symmetric metric.

In section 3 we apply these results to the Schwarzschild and cosmological
FRW metrics, obtain the general Penrose limits and discuss in some
detail the emergence of the $u^{-2}$ singular HPW behaviour in their
near-singularity limits. The FRW metrics in particular, with the freedom
in specifying their perfect-fluid equation of state, will allow us to
anticipate some of the features that will then reappear in the general
discussion of section 4.

In section 4 we introduce the Szekeres-Iyer metrics (section 4.1),
analyse their null geodesics (section 4.2) and their Penrose limits
(sections 4.3). In section 4.4 we supplement this by an analysis of the
DEC in these models and prove the two statements made above. Finally,
section 5 contains various comments on applications of these results
to string theory, open questions and future work.

Even though somewhat outside the main line of this paper, in Appendix
A we elaborate, following the suggestion of \cite{patricot}, how to
use solutions of the Hamilton-Jacobi equations and their corresponding
null geodesic congruences to construct adapted coordinates which are
the starting point of the more traditional approach \cite{penrose,bfp}
to Penrose limits. Appendix B summarises some minor variations of the
calculations of section 2.3, and in Appendix C we list the non-vanishing
components of the Ricci and Einstein tensors for the near-singularity
Szekeres-Iyer metrics.

\section{Penrose Limits via  Geodesic Deviation}

In \cite{bbop1} it was shown that the wave profile $A_{ab}(u)$ of the
Penrose limit plane wave metric associated to a null geodesic $\gamma$
in a space-time with metric $g_{\mu\nu}$ 
can be obtained directly from the curvature tensor of the original metric,
\be
A_{ab}(u) = -R_{a+b+}|_{\gamma}\;\;.
\label{keyeq}
\ee
Here the components refer to a parallel pseudo-orthonormal frame along
$\gamma$,
\be
ds^2|_{\gamma} = 2 E^+ E^- + \d_{ab}E^a E^b
\label{pf1}
\ee
with $E_+\equiv \del_u$ the tangential direction.

This relation is reminiscent of, but should not be
confused with, the well-known relation
\be
\bar{R}_{aubu} = - A_{ab}(u) 
\ee
expressing the sole non-vanishing curvature component of the plane
wave metric 
\be
d\bar{s}^2= 2dudv + A_{ab}(u)x^a x^b du^2 + d\vec{x}^2
\ee
in terms of $A_{ab}(u)$. Indeed, the key observation of \cite{bbop1}
in this respect was that these curvature components are directly related
to those of the original (pre-Penrose limit) metric via (\ref{keyeq}).

Equivalently, $A_{ab}(u)$ can be characterised as the
transverse null geodesic deviation matrix \cite[Section 4.2]{HE} 
of the original metric,
\be
\frac{d^2}{du^2} Z^a = A_{ab}(u) Z^b\;\;,
\label{gde}
\ee
with $Z$ the transverse geodesic deviation vector. 

The equivalence of (\ref{gde}) and the characterisation (\ref{keyeq}) of
$A_{ab}(u)$ obtained in \cite{bbop1} is a standard result in the theory
of null congruences (essentially the Raychaudhuri equation). We have
found that in practice this is not only a geometrically transparent but
frequently also a calculationally efficient way of determining the wave
profile $A_{ab}(u)$, and for this reason we will explain this procedure
in some detail in this section.

\subsection{The Penrose Limit and the Null Geodesic Deviation Equation}

To establish the relation between (\ref{keyeq}) and (\ref{gde}),
we embed the null geodesic
$\gamma$ into some (arbitrary) null geodesic congruence.
Via parallel transport one can construct a parallel pseudo-orthonormal
frame $E^A$, $A=+,-,a$, along the null geodesic congruence,
\be 
ds^2= 2E^+E^- + \d_{ab}E^aE^b\;\;,\;\;\;\;\;\;\nabla_u E^A=0
\label{ppof}
\ee
such that the component $E_+$ of the co-frame $E_A$ is
\be
E_+=\dot{x}^{\mu}\del_{\mu}~,~~~~~~E_+|_\gamma=\del_u \;\;,
\ee
i.e. the restriction of $E_+$ to every null geodesic is the tangent
vector of the null geodesic.

Infinitesimally the congruence is
characterised by the {\em connecting vectors} $Z$ representing
the separation of corresponding points on neighbouring curves and
satisfying the equation
\be
L_{E_+} Z=[E_+,Z]=  \nabla_{E_+} Z -
\nabla_Z E_+ = 0\;\;. \label{aplie}
\ee
In a parallel frame, covariant derivatives along the congruence become
partial derivatives, 
\be
E^\mu_+\nabla_\mu(Z^A E_A)= \nabla_u(Z^A E_A) = (\del_u Z^A)E_A\;\;,
\ee
and since $E_+$ is null one has
\be
g(E_+,E_+)=0
\Ra (\nabla_{A} E_+)^{-} = 0\;\;. \label{minus}
\ee
Hence
(\ref{aplie}) implies that $(d/du) Z^- =0$, and
we can set $Z^-=0$ without loss of generality. Then, using the geodesic
equation $\nabla_uE_+ = 0$, one finds that
\be
\nabla_Z E_+
= Z^b (\nabla_b E_+)^a E_a + Z^b (\nabla_b E_+)^{+}E_+
\ee
and the connecting vector equation (\ref{aplie}) becomes
\be
\frac{d}{du}Z^a = B^{a}_{\;b}Z^b \label{duz}
\ee
with
\be
B^{a}_{\;b} = (\nabla_b E_+)^a \equiv E^a_{\nu} E^{\mu}_{b}
\nabla_{\mu}E_+^{\nu}\;\;,
\label{bab}
\ee
and $Z^+$ determined by the
$Z^a$ via
\be
\frac{d}{du}Z^+ = Z^b(\nabla_b E_+)^+\;\;.
\ee
For later use we note that (\ref{minus}) implies that the trace of $B$ is
\be
\tr B \equiv B^{a}_{\;a} = \nabla_{\mu}E_+^\mu = \frac{1}{\sqrt{-g}}
\partial_\mu(\sqrt{-g}\dot{x}^\mu)\;\;\label{trb} \;\;,
\ee
explaining the ubiquity of logarithmic derivatives in the examples to be
discussed below.

It follows from (\ref{duz})
that the transverse components $Z^a$ satisfy the null geodesic
deviation equation
\be
\frac{d^2}{du^2} Z^a = A_{ab}(u) Z^b\;\;.
\label{apgde}
\ee
where
\be
A^{a}_{\; b} = \frac{d}{du} B^{a}_{\;b}+ B^{a}_{\;c} B^{c}_{\;b}\;\;.
\ee
Note that (\ref{apgde}) is just a
(time-dependent) harmonic oscillator equation with $(-A_{ab}(u))$ the
matrix of frequency squares.

A routine calculation now shows that
\be
A^{a}_{\;b} = E^a_{\nu} E^{\mu}_{b}
R^{\nu}_{\;\lambda\rho\mu}\dot{x}^{\lambda}\dot{x}^{\rho} = -R^a_{\;+b+}\;\;,
\label{apk}
\ee
with $R$ the Riemann curvature tensor of the metric $g$,
establishing the equivalence of (\ref{keyeq}) and (\ref{gde}).

Alternatively, this can be understood in terms of the standard evolution
equations for the expansion, shear and twist of a null geodesic congruence
(see e.g.\ \cite[Section 4.2]{HE} or \cite[Section 9.2]{Wald}),
which are equal to the trace, trace-free symmetric and anti-symmetric
part of $B_{ab}$,
\be
B_{ab} = E_{a}^{\nu}E_{b}^{\mu}\nabla_{\mu}p_{\nu}\;\;,
\;\;\;\;\;\;p_{\nu} = g_{\nu\lambda}\dot{x}^{\lambda}\;\;,
\ee
respectively. From this point of view, the symmetry
of $A_{ab}$, i.e.\ the vanishing of the antisymmetric part of $\dot{B}
+ B^2$, is equivalent to the evolution equation for the twist, and
the equivalence of (\ref{keyeq}) and (\ref{gde}) is the content of the
evolution equation for the symmetric part of $B_{ab}$ whose trace
is the Raychaudhuri equation for null geodesics.

We see from (\ref{apk}) that, even though $B_{ab}$ depends on the
properties of the null geodesic congruence, the particular combination of
expansion, shear and twist and their derivatives appearing in $A_{ab}$
depends only on the components of the curvature tensor and the parallel
frame along the original null geodesic. In particular, the geodesic
deviation matrix $A_{ab}(u)$ is independent of how the null geodesic
$\gamma$ is embedded into some null congruence.

\subsection{Penrose Limits, Geodesic Congruences and Hamilton-Jacobi
Equations}

Using the geodesic deviation approach to calculate Penrose limits, as
outlined above, is obviously a geometrically transparent and appealing
way of interpreting the Penrose Limit and determining $A_{ab}(u)$. It
is somewhat less economical (economical in the sense of introducing
the least amount of additional structure) than the characterisation
(\ref{keyeq}) of $A_{ab}(u)$ in terms of the Riemann tensor, which
only requires a parallel frame along the original null geodesic and
not an entire geodesic congruence. However, it may nevertheless be a
calculationally more efficient approach if one is in a situation where one
has a natural candidate geodesic congruence (so that one does not have
to construct one first). In this case, the calculation of $A_{ab}(u)$
via geodesic deviation provides a shortcut to the calculation of the relevant
components of the Riemann tensor.

Both these covariant characterisations of the Penrose Limit are certainly
more elegant than the standard systematic aproach to determining
Penrose Limits \cite{penrose,gueven,bfp} which not only relies on the
existence of some special (twist-free) null geodesic congruence, but
also requires other auxiliary constructs like Penrose coordinates (i.e.\
coordinates adapted to the congruence) and the coordinate transformation
from Rosen to Brinkmann coordinates. Nevertheless, this is still frequently
a useful way of performing calculations, in particular when combined with
the systematic Hamilton-Jacobi approach to constructing adapted coordinates
first pointed out in \cite{patricot}, and for this reason we provide a more
detailed account of this construction in Appendix A.

In practice, therefore, the geodesic deviation approach is useful
if there is a natural geodesic congruence. Such a null geodesic congruence
can be easily constructed whenever one has a solution to the Hamilton-Jacobi
equation 
\be
g^{\mu\nu} \partial_\mu S \partial_\nu S=0
\ee
for null geodesics. Indeed, setting
\be
\dot{x}^{\mu}=g^{\mu\nu}\del_{\nu}S\;\;, 
\ee
one obviously has
\be
\dot{x}^\rho\nabla_\rho \dot{x}^\mu= g^{\rho\sigma}
g^{\mu\nu}\nabla_\rho\del_{\nu}S \del_{\sigma}S
=\frac{1}{2} g^{\mu\nu} \partial_\nu (g^{\rho\sigma}
\del_\rho S \del_\sigma S)=0\;\;,
\label{sg}
\ee
so that this defines a null geodesic congruence.

In particular, whenever the Hamilton-Jacobi equation can be separated
(this includes all space-times investigated in this paper) the null
geodesic equations become first order and the natural null geodesic
congruence is parameterised by the integration constants of these first
order equations.

For a geodesic congruence defined by a solution to the Hamilton-Jacobi
equation, the equation for $B_{ab}$ (\ref{bab}), is
\be
B_{ab}=E_a^\mu E_b^\nu \nabla_\mu \partial_\nu S
\label{hjbab}
\ee
and the equation for the trace of $B$, (\ref{trb}), is
\be
B^a{}_a=\nabla^\mu\partial_\mu S\;\;.
\label{hjtrb}
\ee
Therefore $B$ is the covariant Hessian of the HJ function $S$ evaluated
in a  parallel frame and the trace of $B$ is the Laplacian of the
Hamilton-Jacobi function with respect to the space-time metric. Since
$B$ is a symmetric matrix, the corresponding null geodesic congruence
is twist free.

\subsection{The Penrose Limits of a Static Spherically Symmetric Metric}

To illustrate the geodesic deviation approach to Penrose limits,
and as a preparation for the calculations of sections 3 and 4,
we now show how to quickly determine all the Penrose limits of a
static spherically symmetric metric. We start with the metric in
Schwarzschild-like coordinates (the extension to isotropic coordinates,
brane-like metrics with extended world volumes, or null metrics
is straightforward and is discussed in Appendix B),
\bea
ds^2  &=& -f(r) dt^2 + g(r) dr^2 + r^2 d\Omega_d^2 \non
d\Omega_d^2 &=& d\theta^2 + \sin^2 \theta d\Omega_{d-1}^2 \;\;.
\eea
Taking the Penrose limit entails first choosing a null geodesic. Because
of the rotational symmetry in the transverse direction,
without loss of generality we can choose
the null geodesic to lie in the $(t,r,\theta)$-plane.
The symmetries reduce the geodesic equations to the first integrals
\bea
\dot{t} &=& E/f(r)\non
\dot{\theta} &=& L/r^2\non
\dot{r}^2 &=& E^2/f(r)g(r) - L^2/ g(r)r^2\;\;,
\eea
where $E$ and $L$ are the conserved energy and angular momentum respectively.
This defines a natural geodesic congruence, corresponding to the
Hamilton-Jacobi function
\be
S=-E t+ L \theta+ R(r)
\ee
with 
\be
(\frac{d}{dr} R)^2=g f^{-1} E^2- r^{-2} g L^2
\ee
and allows us to calculate $B_{ab}$. 

We first construct the parallel frame. We have
\be
E_+ = \dot{r}\del_r + \dot{t}\del_t + \dot{\theta}\del_\theta\;\;,
\;\;\;\;\;\; E_+|_\gamma=\del_u\;\;,
\ee
and we will not need to be more specific about $E_-$.
The transverse components are $E_a=(E_1,E_{\hat{a}})$, with
$\hat{a}=2,\ldots,d$ referring to the transverse $(d-1)$-sphere. Since
there is no evolution in these directions, the $E_{\hat{a}}$ are the obvious
parallel frame components
\be
E_{\hat{a}} = \frac{1}{r\sin\theta} e_{\hat{a}}
\ee
with $e_{\hat{a}}$ an orthonormal coframe for $d\Omega_{d-1}^2$.
The transverse $SO(d)$-symmetry implies
\bea
B_{1\hat{a}}&=&A_{1\hat{a}}=0\non
B_{\hat{a}\hat{b}}(u) &=&B(u)\delta_{\hat{a}\hat{b}}\non
A_{\hat{a}\hat{b}}(u) &=&A(u)\delta_{\hat{a}\hat{b}}\;\;.
\eea
Moreover, because of (\ref{trb}) we have
\be
B_{11}(u) = \nabla_{\mu}\dot{x}^{\mu}(u) - (d-1) B(u)
\ee
so that we only have to calculate $B_{22}(u) = B(u)$,
for which one finds (with, say, $e_2=\del_\phi$)
\be
B_{22} =\Gamma^\phi_{\phi r}\dot{r} + \Gamma^\phi_{\phi\theta} \dot{\theta}
= \partial_u \log(r(u) \sin\theta(u))\;\;,
\ee
or
\be
B_{\hat{a}\hat{b}}(u) = \delta_{\hat{a}\hat{b}}
\del_u \log (r(u) \sin\theta(u))\;\;.
\label{b22}
\ee
Since
\be
\tr B = \del_u \log \Biggl(\dot{r}r^d \sin^{d-1}\theta\sqrt{f(r)g(r)}\Biggr)
\ee
one finds
\be
B_{11}(u) = 
\del_u \log \Biggl(r(u) \dot{r}(u) \sqrt{f(r(u))g(r(u))}\Biggr)\;\;.
\label{b11}
\ee
Now, in general, for $B_{ab}(u)$ of the logarithmic derivative form
\be
B_{ab}(u) = \delta_{ab}\del_u \log K_a(u)
\label{k1}
\ee
one has
\be
A_{ab}(u) = \delta_{ab} K_a(u)^{-1} \del_u^2 K_a(u)
\label{k2}
\ee
and therefore
\bea
A_{11} &=& (r\dot{r}\sqrt{fg})^{-1}\del_u^2 (r\dot{r}\sqrt{fg})\non
A_{\hat{a}\hat{b}} &=&\delta_{\hat{a}\hat{b}}
(r\sin\theta)^{-1}\del_u^2 (r\sin\theta)\;\;.
\eea
In particular, for the transverse components one has the universal
result
\be
A_{\hat{a}\hat{b}}(u)
=\delta_{\hat{a}\hat{b}}(\frac{\ddot{r}(u)}{r(u)}-\frac{L^2}{r(u)^4})\;\;.
\label{a22}
\ee

\section{Examples}

\subsection{Schwarzschild Plane Waves and their Homogeneous
Near-Singularity Limits}

As a concrete example we will now consider the Penrose limits of the
$D=(d+2)$-dimensional Schwarzschild metric\footnote{The Penrose limit for
certain special (radial $L=0$) null geodesics in the AdS-Schwarzschild
metric has been discussed before e.g.\ in \cite{lzjs,dmlz}. The general
case, using the Hamilton-Jacobi method to construct adapted coordinates, 
was presented in \cite{mbictplec}. For a general discussion of limits of 
the Schwarzschild metric not depending on additional parameters like $L$
see \cite{reboucas}.}
\be
ds^2 = -f(r)dt^2 + f(r)^{-1} dr^2 + r^2 d\Omega_d^2
\label{ssm}
\ee
where
\be
f(r) = 1 - \frac{2m}{r^{d-1}}\;\;.
\label{ssf}
\ee
We are assuming that $D\geq 4$ and note that evidently only for $D=4$
is $m$ the ADM mass of the black hole.

In this case we have
\be
\dot{r}^2 = E^2 - L^2 f(r) r^{-2} \equiv E^2 - 2 V_{eff}(r)\;\;,
\label{rdot2}
\ee
where $V_{eff}(r)$ is the usual effective potential, with respect to which
$r(u)$ satisfies the Newtonian equation of motion
\be
\ddot{r} = - V_{eff}^{\prime}(r)\;\;.
\label{ddotr}
\ee
It follows that
\bea
A_{22}(u) = \ldots = A_{dd}(u) &=&
\frac{\ddot{r}(u)}{r(u)}-\frac{L^2}{r(u)^4} \non
&=& -\frac{(d+1)mL^2}{r(u)^{d+3}}\;\;,
\label{ss1}
\eea
where $r(u)$ is the solution to the geodesic (effective potential)
equation (\ref{rdot2}).
Moreover, since the Schwarzschild metric is a vacuum
solution, this is a vacuum plane wave with $\Tr A(u) = \d^{ab}A_{ab}(u)=0$,
so that
\be
A_{11}(u) = \frac{(d+1)(d-1)mL^2}{r(u)^{d+3}}
\label{ss2}
\;\;.
\ee

There are a number of facts that can be readily deduced from this result:
\begin{itemize}
\item First of all, we see that the Penrose limit of the
Schwarzschild metric is flat for radial null geodesics, $L=0$. We
could have anticipated this on general grounds because in this case
the setting is $SO(d+1)$-invariant, implying $A_{ab}(u)\sim \d_{ab}$,
which is incompatible with $\Tr A =0$ unless $A_{ab}(u)=0$. This should,
however, not be interpreted as saying that the radial Penrose limit
of the Schwarzschild metric is Minkowski space. Rather, the space-time
``ends'' at the value of $u$ at which $r(u)=0$, say at $u=0$. Perhaps
the best way of thinking of this metric is as a time-dependent orbifold
of the kind studied recently in the context of string cosmology (see e.g.\
\cite{lcmc} and references therein).
\item
We also learn that the Penrose limit is a symmetric plane wave
($u$-independent wave profile) if $r(u)=r_*$ is a null geodesic
at constant $r$. Setting $\ddot{r}=\dot{r}=0$, one finds that
\be
r_*^{d-1} = (d+1)m
\ee
(the familiar $r=3m$ photon orbit for $D=4$), with the constraint
\be
r_*^2 = \frac{d-1}{d+1}\frac{L^2}{E^2}
\label{constraint}
\ee
on the ratio $L/E$. Precisely because they lead to symmetric plane waves,
with a well-understood string theory quantisation, such constant $r$
Penrose limits have attracted some interest in the literature.
\item
Moreover we see that the resulting plane wave metric for $L\neq 0$
is singular iff
the original null geodesic runs into the singularity, which will happen
for sufficiently small values of $L/E$.
\end{itemize}

We will now take a closer look at the $u$-dependence of the wave profile
near the singularity $r(u)=0$.  We thus consider sufficiently small
values of $L/E$ in order to avoid the angular momentum barrier.

For small values of $r$, the dominant term in the differential equation
(\ref{rdot2}) for $r$ is (unless $L =0$, a case we already dealt with above)
\be
\dot{r} = \sqrt{2m}L r^{-(d+1)/2}\;\;.
\ee
This implies that
\be
r(u)^{d+3} = \frac{m L^2(d+3)^2}{2}u^2\;\;.
\ee
Thus the behaviour of the Penrose limit of the Schwarzschild metric as
$r\ra 0$ is
\be
A_{11}(u) = - \omega'^{2}_{SS}(d)u^{-2}
\ee
and
\be
A_{22}(u)=\ldots = A_{dd}(u) = -\omega^{2}_{SS}(d)u^{-2},
\ee
with frequencies
\be
\omega'^{2}_{SS}(d)= -\frac{2(d^2-1)}{(d+3)^2}\;\;.
\ee
and
\be
\omega^{2}_{SS}(d)= \frac{2(d+1)}{(d+3)^2}\;\;.
\ee
We note the following:
\begin{itemize}
\item First of all,
in this limit one finds a singular homogeneous plane wave of the type
(\ref{2}). As we will see later, this scale invariance of the near-singularity
Penrose limit can be attributed to the power-law scaling behaviour of the
near-singularity Schwarzschild metric.
\item
Moreover, the dependence on $L$ and $m$ has dropped out.
The metric thus exhibits a universal
behaviour near the singularity which depends only on the space-time
dimension $D=d+2$, but neither on the mass of the black hole nor
on the angular momentum of the null geodesic used to approach the
singularity. For example, for $D=4$ one has
\be
\omega_{SS}^2(d=2)=\frac{6}{25} \;\;.
\ee
\item
The frequencies are bounded by
\be
\omega'^{2}_{SS}(d) < 0 < 
\omega^{2}_{SS}(d) < \frac{1}{4}\;\;.
\ee
\item
Finally, we note that the above result is also valid for 
(A)dS black holes since the presence of a cosmological constant is irrelevant
close to the singularity.
\end{itemize}

\subsection{FRW Plane Waves and their Homogeneous Near-Singularity Limits}

As another example 
we consider the Penrose limit of the $D=(n+1)$-dimensional FRW metric
\be
ds^2 = -dt^2 + a(t)^2(dr^2 + f_k(r)^2 d\Omega_{n-1}^2)\;\;,
\ee
where $f_k(r)=r,\sin r,\sinh r$ for $k=0,+1,-1$ respectively.

Since the spatial slices are maximally symmetric, up to isometries
there is a unique null geodesic and hence a unique Penrose limit.
So without loss of generality we shall consider null geodesics which
have vanishing angular momentum on the transverse sphere.

Then, with a suitable scaling of the affine parameter,
the null geodesic equations can be written as
\be
\frac{d}{du} t(u) = \pm a(t(u))^{-1}\;\;,\;\;\;\;\;\;
\frac{d}{du} r(u) = a(t(u))^{-2}
\label{ngc}
\ee
(and in what follows, we choose the upper sign in the first equation).
Thus
\be
E_+=\del_u = a^{-1}\del_t + a^{-2} \del_r\;\;,
\ee
and this can be extended to a parallel pseudo-orthonormal frame by
\bea
E_- &=& \frac{1}{2}(-a\del_t + \del_r)\non
E_a &=& (af_k)^{-1} \hat{e}_a\;\;,
\eea
where $\hat{e}_a$ is an orthonormal frame for $d\Omega_d^2$, $d=n-1$.

The transverse rotational symmetry implies that
$B_{ab}(u)=B(u) \d_{ab}$ and $A_{ab}(u)=A(u)\d_{ab}$. Therefore, 
to determine $B(u)$ it suffices to compute the trace of $B_{ab}(u)$, 
\be
\tr B =\partial_u \log\biggl(a^{n-1} f_k^{n-1}\biggr)\;\;,
\ee
implying
\be
B(u)=\partial_u \log\bigl(a f_k\bigr)\;\;.
\ee
Using $\frac {d^2}{dr^2} f_k=-k f_k$, one finds
\be
A_{ab}(u) = \d_{ab}A(u)= \d_{ab}
\left(\frac{\ddot{a}(u)}{a(u)}-\frac{k}{a(u)^4}\right)\;\;.
\label{frwa}
\ee
This is the precise analogue of the expression (\ref{a22}) obtained in
the static spherically symmetric case, the spatial curvature $k$ now playing 
the role of the angular momentum $L^2$.

This can now be rewritten in a variety of ways to obtain insight into the
properties of this FRW plane wave. For example, writing this in terms of
$t$-derivatives (in order to make use of the Friedmann equations), we find
\be
A(u(t)) =
\frac{1}{a(t)^2}(\frac{a''(t)}{a(t)}-
\frac{k+a'(t)^2}{a(t)^2})\;\;.
\label{frw1}
\ee
where $a(t)$ is determined by the Einstein (Friedmann)
equations, $u(t)$ by $du = a(t) dt$, and $a'=\frac{d}{dt} a$.
The Friedmann equations
\bea
\frac{a'(t)^2 + k}{a(t)^2} &=& \frac{16\pi G}{n(n-1)} \rho(t)\non
\frac{a''(t)}{a(t)} &=& - \frac{8\pi G}{n(n-1)}[(n-2)\rho(t) +n P(t)]
\;\;,
\eea
imply
\be
\frac{a'(t)^2 + k}{a(t)^2}
- \frac{a''(t)}{a(t)} = \frac{8\pi G}{(n-1)}[\rho(t) + P(t)]
\;\;,
\ee
so that one finds that the wave profile of the FRW plane wave can be written
compactly as
\be
A(u) = -\frac{8\pi G}{n-1} \frac{\rho(u) + P(u)}{a(u)^2}\;\;.
\label{frw2}
\ee
One immediate consequence is that the Penrose limit is flat if and only if
$\rho + P =0$, corresponding to having as the only matter content a
cosmological constant. This is in agreement with the result \cite{bfp} that
every Penrose limit of a maximally symmetric space-time is flat.

We will now study the behaviour of $A(u)$ near a singularity, and to be
specific we choose the usual equation of state
\be
P(t)=w\rho(t)\;\;.
\ee
We consider $w > -1$ ($w= -1$ would correspond to the case $\rho + P =0$ 
already dealt with above) and introduce the positive parameter
\be
h(n,w) = \frac{2}{n(1+w)}
\ee
and the positive constant (constant by the continuity equation for $\rho$)
\be
C_h = \frac{16\pi G}{n(n-1)}\rho(t)a(t)^{2/h}\;\;,
\ee
in terms of which the Friedmann equations read
\bea
a'(t)^2 &=& C_h a(t)^{(2h-2)/h} - k \label{freq1}\\
a''(t) &=& \trac{h-1}{h}C_h a(t)^{(h-2)/h} \label{freq2}\;\;.
\eea
Thus the universe is decelerating for $0<h<1$ and accelerating for $h>1$,
the critical case $h=1$ corresponding to $w_c = -1 + 2/n$ (the familiar
dark energy threshold $w_c=-1/3$ for $n=3$).

We first consider the case $k=0$. In that case one has
\be
a(t) \sim t^h\;\;,
\ee
and therefore
\be
a(u) \sim u^{h/h+1}\;\;.
\label{frwau}
\ee
It then follows immediately from (\ref{frwa}) 
that, more explicitly, the $u$-dependence of $A(u)$
is\footnote{This generalises the result reported in \cite{bfp}.}
\be
A(u) = -\omega^2_{FRW}(h,k=0) u^{-2}\;\;,
\label{Au}
\ee
where
\be
\omega^2_{FRW}(h,k=0) = \frac{h}{(1+h)^2}\;\;.
\label{frwomega}
\ee
We see that the Penrose limit of a spatially flat FRW universe with
equation of state $P=w\rho$ is exactly a singular homogeneous plane wave
of the type (\ref{2}). 

The frequency square $\omega^2_{FRW}(h,k=0)$ has the following properties: 
\begin{itemize}
\item
Since
\be
\omega^2_{FRW}(h,k=0) = \omega^2_{FRW}(1/h,k=0)\;\;,
\ee
for every accelerating (inflating) solution of the $k=0$
Friedmann equations there is precisely one decelerating solution with the
same Penrose limit. The self-dual point $h=1$ corresponds to the linear
time-evolution $a(t)\sim t$.
\item The frequency squares are again bounded by
\be
\omega^2_{FRW}(h,k=0) \leq \frac{1}{4}\;\;,
\ee
with equality attained for $h=1$.
\item
Curiously, the frequencies obtained in the Penrose limit of the
Schwarzschild metric (in all but one of the directions) are precisely
those of a dust-filled FRW universe, $P=w=0$, of the same dimension
$n=d+1$,
\be
\omega^{2}_{SS}(d) = \omega^{2}_{FRW}(h,k=0)\;\;,
\ee
e.g.\ $6/25$ for $n=3$.
\end{itemize}

It is clear that for $k=0$, when only the first term in (\ref{frwa})
is present, this homogeneous $u^{-2}$-behaviour is a consequence of the
exact power-law behaviour of $a(t)$ and hence $a(u)$. Let us now consider
what happens for $k\neq 0$, when there is a competition between the
two terms in (\ref{frwa}) as one approaches the singularity. 

One might like to argue that, even for $k\neq 0$, one finds the same
behaviour provided that the matter term dominates over the curvature
term in the Friedmann equation (\ref{freq1}) as $a \ra 0$. This
happens for $0<h<1$, and this argument is correct as one can also see
that in this range the first term in (\ref{frwa}), proportional to
$u^{-2}$, indeed dominates over the second (curvature) term which (cf.\
(\ref{frwau})) is proportional to $u^{-4h/(h+1)}$. Thus for $0<h<1$
the near-singularity limit of the FRW plane wave is a homogeneous plane
wave with $k$-independent frequencies (\ref{frwomega}),
\be
0<h<1:\;\;\;\;
\omega^2_{FRW}(h,k) = \omega^2_{FRW}(h,k=0) = \frac{h}{(1+h)^2}\;\;.
\ee

Now let us look at what happens as one passes from a decelerating to
a critical ($h=1$) and then accelerating ($h>1$) universe. First of all,
for $h=1$, both terms on the right hand side of the Friedmann equation
(\ref{freq1}) contribute equally (they are constant), and correspondingly
both terms in (\ref{frwa}) are proportional to $u^{-2}$. Thus one finds
a homogeneous plane wave, but with a curvature-induced shift of the
frequency,
\be
h=1:\;\;\;\;
\omega^2_{FRW}(h=1,k) = \omega^2_{FRW}(h=1,k=0) + k c^2 = \frac{1}{4} + kc^2
\ee
for some constant $c$. In particular, in the spatially closed case $k=+1$
(this requires $C_h > 1$), one now finds frequency squares that are
larger than $1/4$.  This is a borderline behaviour in the sense that,
as can easily be seen from (\ref{freq1}), the initial singularity for
$k=+1$ ceases to exist for $h>1$.

It thus remains to discuss the case $k=-1$ and $h>1$. Given
the previous discussion, one might be tempted to think that now the
second term in (\ref{frwa}) will dominate over the first, leading to a
non-homogeneous and more singular $u^{-4h/(h+1)}$-behaviour. This is,
however, not the case, as (\ref{frwau}) now represents the leading
behaviour at large $a(u)$. At small $a(u)$, the leading behaviour is,
exactly as for $h=1$, determined by the constant curvature term in
(\ref{freq1}). Thus even in this case one finds a singular homogeneous
plane wave, with frequency
\be
h>1:\;\;\;\; 
\omega^2_{FRW}(h,k=-1) = \frac{1}{4} - c^2
\ee
once again bounded from above by $1/4$.

\section{The Universality of Penrose Limits of Power-Law Type Singularities}

In the previous section we have presented some evidence for a
remarkable
\begin{quote}
{\bf Conjecture:} Penrose limits of physically reasonable space-time
singularities are singular homogeneous plane waves with wave profile
$A_{ab}(u) \sim u^{-2}$.
\end{quote}
In this section we will show how to prove this conjecture for a very
large class of  physical singularities of spherically symmetric type. We
will in the process also see some examples of ``extreme'' stress-energy
tensors that give rise to a different behaviour.

\subsection{Szekeres-Iyer Metrics}

The homogeneity of the Penrose limit (geodesic deviation) that we have
found in the above examples appears to reflect a power-law scaling
behaviour of the metric near the singularity.  Thus to assess the
generality of this kind of result, one needs to enquire about the
generality of space-time singularities exhibiting such a power-law
behaviour.

In \cite{SI} (see also \cite{CS}), in the context of investigations of the
Cosmic Censorship Hypothesis, Szekeres and Iyer studied a large class of
four-dimensional spherically symmetric metrics they dubbed ``metrics with
power-law type singularities''. Such metrics encompass practically all
known singular spherically symmetric solutions of the Einstein equations,
in particular all the FRW metrics, Lema\^itre-Tolman-Bondi dust solutions,
cosmological singularities of the Lifshitz-Khalatnikov type, as well as
other types of metrics with null singularities.

In ``double-null form'', these metrics (in $d+2$ dimensions) take the form
\be
ds^2 = -\ex{A(U,V)}dUdV + \ex{B(U,V)} d\Omega_d^2\;\;,
\label{sim}
\ee
where $A(U,V)$ and $B(U,V)$ have expansions
\bea
A(U,V)&=& p \ln x(U,V) + \mathrm{regular\; terms}\non
B(U,V)&=& q \ln x(U,V) + \mathrm{regular\; terms}
\eea
near the singularity surface $x(U,V)=0$. 

Generically, the residual coordinate transformations $U\ra U'(U)$, $V\ra V'(V)$
preserving the form of the metric (\ref{sim}) can be used to make
$x(U,V)$ linear in $U$ and $V$,
\be
x(U,V) = kU+lV\;\;,\;\;\;\;\;\;k,l=\pm 1, 0\;\;,
\ee
with $\eta = kl = 1, 0, -1$ corresponding to spacelike, null and timelike
singularities respectively.
This choice of gauge essentially fixes the coordinates
uniquely, and thus the ``critical exponents'' $p$ and $q$ contain
diffeomorphism invariant information. 

The Schwarzschild metric, for example, has $p=(1-d)/d$ and $q=2/d$, as is
readily seen by starting with the metric in Eddington-Finkelstein or
Kruskal-Szekeres coordinates and transforming to the Szekeres-Iyer gauge.
Alternatively, if one is just interested in the leading behaviour of
the metric, one can simply expand the metric near $r=0$ and then go 
to tortoise-coordinates. 

We will focus on the behaviour of these geometries near the singularity
at $x=0$, where the metric is
\be
ds^2 = - x^p dU dV + x^q d\Omega_d^2\;\;.
\label{leading}
\ee
For generic situations this leading behaviour is sufficient to discuss the
physics near the singularity. In certain special cases, for particular
values of $p,q$ or for null singularities, this leading behaviour
cancels in certain components of the Einstein tensor and the subleading
terms in the above metric become important for a full analysis of the
singularities \cite{SI,CS}. The analysis then becomes more subtle
and we will not discuss these cases here. In the following we will
consider exclusively the metric (\ref{leading}) which, for $\eta\neq 0$
and generic values of $p$ and $q$, captures the dominant behaviour of
the physics near the singularity.

For $\eta \neq 0$ we define $y= kU - lV$ and choose $k=\eta l=1$. Then the
metric takes the form
\be
ds^2 = \eta x^p dy^2 - \eta x^p dx^2 + x^q d\Omega_d^2 \;\;.
\label{sim1}
\ee
With the further definition $r=x^{q/2}$ (for $q \neq 0$), this has the
standard form of a spherically symmetric metric. We will come back to
this below in order to be able to make direct use of the analysis of
section 2.3.

For $\eta = 0 $, on the other hand, we choose $x=U$, $y=-V$, so that the
metric is 
\be
ds^2 = x^p dx dy + x^q d\Omega_d^2\;\;,
\label{nullmetric}
\ee
which has the form of the spherically symmetric null metrics 
analysed in Appendix B.

\subsection{Null Geodesics of Szekeres-Iyer Metrics}

In terms of the conserved momenta $P$ and $L$ associated with $y$ and,
say, the colatitude $\theta$ of the $d$-sphere, 
\be
d\Omega_d^2 = d\theta^2 + \sin^2 \theta d\Omega_{d-1}^2\;\;,
\ee
in particular
\be
x^q \dot{\theta} = L\;\;,
\ee
the null geodesic condition (for any $\eta$) is equivalent to
\be
\dot{x}^2 = P^2 x^{-2p} + \eta L^2 x^{-p-q}\;\;,
\label{sig}
\ee
To understand the null geodesics near $x=0$, we begin by extracting
as much information as possible from this equation,
recalling that due to the expansion around $x=0$ we can only trust the
leading behaviour of this equation as $x\rightarrow 0$.

Unless $p=q$, one of the two terms on the right-hand-side of
(\ref{sig}) will dominate as $x\ra 0$, and thus the generic behaviour of
a null geodesic near $x=0$ is identical to that of a geodesic with either
$L=0$ or $P=0$. In the former case, one finds
\bea
\mbox{Behaviour 1:} && x(u) \sim u^{1/(p+1)}
\eea
unless $p=-1$ when $x(u) \sim \exp{u}$.  We are only interested
in those geodesics which run into the singularity at $x=0$ at finite
$u$. This happens only for $p>-1$.  In the latter case, corresponding to
null geodesics which asymptotically, as $x \ra 0$, behave like geodesics
with $P=0$, we evidently need $\eta=+1$ (a spacelike singularity),
which leads to
\bea
\mbox{Behaviour 2:} && x(u) \sim u^{2/(p+q+2)}
\eea
unless $p+q = -2$ which again leads to an exponential behaviour. These null
geodesics run into the singularity at finite $u$ for $p+q > -2$.

For $\eta =+1$, the situation regarding null geodesics that reach the
singularity at finite $u$ is summarised in the following table. 

\begin{equation}
\begin{array}{|c|c|c|}\hline
\mathrm{Conditions \; on\;} (P,L) & \mathrm{Constraints \; on\;} (p,q) & \mathrm{Behaviour} \\ \hline
 P\neq 0, L = 0 & p > -1 & 1\\
 P=0, L\neq 0 & p+q > -2 & 2 \\
 P\neq 0, L \neq 0 & p>q, p >-1 & 1 \\
 P\neq 0, L \neq 0 & p<q, p+q >-2 & 2 \\
 P\neq 0, L \neq 0 & p=q>-1 & 1=2\\ \hline 
\end{array}
\end{equation}

For $\eta =-1$, the situation is largely analogous, the main difference
being that now the second term in (\ref{sig}) acts as an angular
momentum barrier preventing e.g.\ geodesics with $L \neq 0$ for $q>p$
from reaching the singularity at $x=0$. These cases are indicated by a
`$-$' in the table below. For the same reason, for $p=q$ one finds the
constraint $|P|>|L|$.

\begin{equation}
\begin{array}{|c|c|c|}\hline
\mathrm{Conditions \; on\;} (P,L) & \mathrm{Constraints \; on\;} (p,q) & \mathrm{Behaviour} \\\hline
 P\neq 0, L = 0 & p > -1 & 1\\
 P=0, L\neq 0 & & - \\
 P\neq 0, L \neq 0 & p>q, p >-1 & 1 \\
 P\neq 0, L \neq 0 & p<q & - \\
 |P|>|L|  & p=q>-1 & 1=2 \\ \hline
\end{array}
\end{equation}

Finally, for $\eta=0$ we find Behaviour 1 for all values of $p$ and
$q$, with the corresponding constraint $p>-1$. 

\subsection{Penrose Limits of Power-Law Type Singularities}

We will now determine the Penrose limits of the Szekeres-Iyer metrics
along the null geodesics reaching the singularity $x=0$ at finite $u$.

For $\eta \neq 0$ we notice that the metric is simply a
special case of a spherically symmetric metric and thus
can be treated using the analysis of section 2.3. Indeed,
with $t=y$ and $r= x^{q/2}$ ($q\neq 0$), 
the metric (\ref{sim1}) takes the form
\bea
ds^2 &=& \eta x^p dy^2 - \eta x^p dx^2 + x^q d\Omega^2\\
&=& \eta r^{2p/q} dt^2 - \frac{4\eta}{q^2}r^{2(p-q+2)/q}dr^2 + r^2 d\Omega^2
\eea
where in the second line the notation of $t$ and $r$ is adapted to 
the case of $\eta = -1$ where the singularity is timelike and $t$ 
is time. We will continue to use this notation even for spacelike 
singularities where $t$ is actually spacelike. 

The case $q=0$ is special, but actually corresponds to a shell crossing
singularity \cite{SI} which is usually not considered to be a true
singularity as the transverse sphere is of constant radius $x^q=1$. Such
singularities arise for instance for certain collisions of spherical dust
shells. From here on we will only discuss $q\neq 0$.

Referring to section 2.3 where such a spherically symmetric metric was 
treated, we can identify
\bea
f(r) &=& -\eta r^{2p/q}\\
g(r) &=& -\frac{4\eta}{q^2}r^{2(p-q+2)/q}.
\eea
We can now appeal to (\ref{b11}, \ref{b22}) to deduce that
\be
B_{ab} = \delta_{ab}\partial_u \log(K_a(u))
\ee
with
\bea
K_1(u) &=& \dot{r}(u)r(u)^{2(p+1)/q}\non
K_2(u) &=& r(u)\sin(\theta(u))
\label{kiu}
\eea
As shown in Appendix B, it is straightforward to extend this analysis
to the case of null singularities, $\eta = 0$, with the result that the
expressions for $K_a(u)$ are identical to those given for $\eta = \pm 1$
in (\ref{kiu}).  We can thus treat all three cases simultaneously.

It follows from the analysis of the previous section that the only 
possibility of interest for $r(u)=x(u)^{q/2}$ is the power-law behaviour
\be
r(u) = u^a\;\;,
\ee
with
\bea
\mbox{Behaviour 1:} & p > -1 & a = q/2(p+1)\\
\mbox{Behaviour 2:} & p+ q > -2 & a = q/(p+q+2)\;\;.
\eea
Clearly, then, $K_1(u)$ is also a simple power of $u$. Specifically
one has (since we are interested in the logarithmic derivatives of
$K_1(u)$, proportionality factors are irrelevant)
\bea
\mbox{Behaviour 1:} && K_1(u) \sim r(u)\non
\mbox{Behaviour 2:} && K_1(u) \sim r(u)^{p/q}\;\;.
\eea
Thus the corresponding component of $A_{ab}(u)$ is
\bea
\mbox{Behaviour 1:} &&
A_{11}(u) = \frac{\ddot{K}_1(u)}{K_1(u)} =  a(a-1)u^{-2}\non
\mbox{Behaviour 2:} &&
A_{11}(u) = \frac{\ddot{K}_1(u)}{K_1(u)} =  
pa/q(pa/q-1)u^{-2}\;\;.
\eea
and the Penrose limit behaves as a singular homogeneous plane wave in this
direction. Since $b(b-1)$ has a minimum $-1/4$ at $b=1/2$, 
this leads to the bound
\be
\omega_1^2 \leq \frac{1}{4}\;\;.
\ee
This is the same range that we found empirically for both the
Schwarzschild and FRW plane waves near the singularity.

The behaviour of $A_{22}$ is more subtle due to the dependence of $K_2(u)$
on $\sin \theta(u)$. The general behaviour is as in (\ref{a22}), namely
\be
A_{22}(u)=\frac{\ddot{r}(u)}{r(u)} - \frac{L^2}{r(u)^4}\;\;.
\label{a222}
\ee
With the power-law behaviour
$r(u) = u^a$, the first term is always proportional to $u^{-2}$.
This term is dominant as $u \ra 0$ when $a < 1/2$, while it is the second
term that dominates for $a > 1/2$ (and leads to a strongly singular
plane wave with profile $\sim u^{-4a}$).
In the special case $a=1/2$, both terms are proportional to $u^{-2}$.
Thus one has, for $L\neq 0$,
\bea
r(u) = u^a & a < \frac{1}{2}: & A_{22}(u) \ra -\omega_2^2 u^{-2}
 \;\;,\;\;\;\;\;\; \omega_2^2 = a(1-a) < \frac{1}{4} \label{a221}\\
 & a = \frac{1}{2}: & A_{22}(u) \ra -\omega_2^2 u^{-2}
 \;\;,\;\;\;\;\;\; \omega_2^2 = \frac{1}{4} + L^2 \geq \frac{1}{4} \\
 & a > \frac{1}{2}: & A_{22}(u) \ra -L^2 u^{-4a}\;\;.
\eea
When $\eta = 0$, Behaviour 1 arises in the entire p-q plane. Thus
$\omega_1^2 \leq 1/4$ and $a$ can take any of the three values above
with the special value $a=1/2$ corresponding to the line $q=p+1$. When
$p\geq q$ and $\eta\neq 0$, $a = q/2(p+1)$ and thus always $a<1/2$. On
the other hand, when $p<q$ and $\eta=1$ we see that $a=q/(p+q+2)$
can take on any value, with $a=1/2$ along the line $q=p+2$ and $a>1/2$
for $q>p+2$.  When $p<q$ and $\eta = -1$ we cannot reach the singularity
along a geodesic with $L\neq 0$.

\begin{figure}
\begin{center}
\includegraphics[width=5 cm]{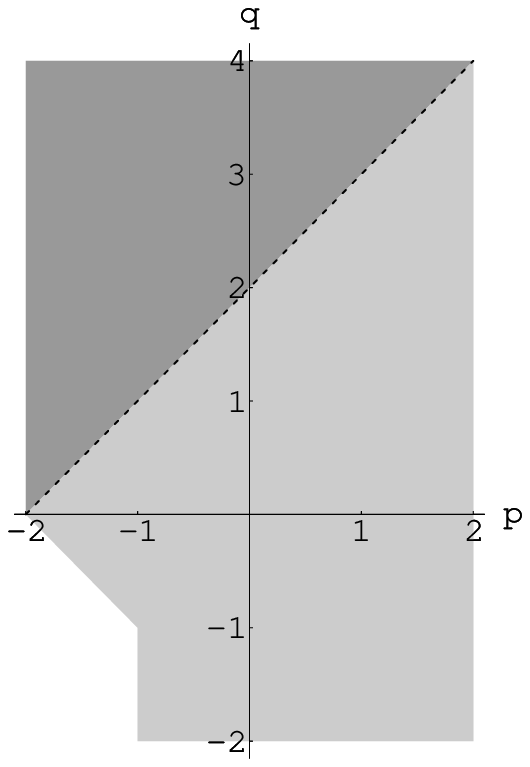}
\hfill
\includegraphics[width=5.1 cm]{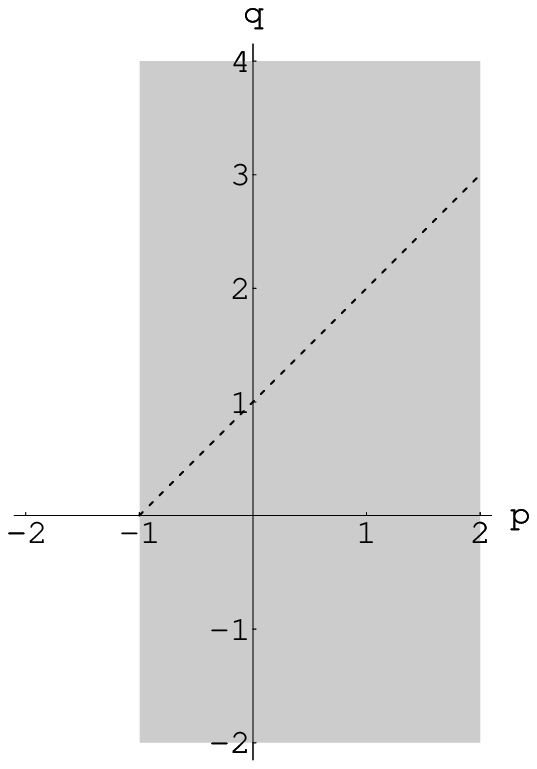}
\end{center}
\caption{The Penrose Limit Phase Diagram in the $p-q$ plane for (a)
spacelike ($\eta=+1$) and (b) timelike ($\eta=-1$) singularities. 
Singular HPWs arise in the light-shaded regions whereas in the 
dark-shaded region there are Penrose limits leading 
to strongly singular (and non-homogeneous) plane waves. (a) The diagram
is bounded on the left by the lines $p=-1$ and $p+q = -2$. The
dashed line $a=1/2 \LRa q=p+2$ separates the two regions, and
only along that line one finds singular HPWs with $\omega_2^2 > 1/4$. 
(b) For $\eta=-1$, one finds singular HPWs with $\omega_2^2 \leq 1/4$ 
for all $(p,q)$ with $p>-1$, $\omega_2^2=1/4$
arising only along the dashed line $a=1/2$ for zero angular momentum,
$L=0$.} 
\end{figure}

When $L=0$, only the first term in (\ref{a222}) is present, and one thus
finds (\ref{a221}) for all values of $a$. Since $L=0$ implies Behaviour 1,
this means $a=q/2(p+1)$. Along the special line $q=2(p+1)$ one has $a=1$
and one finds the ``flat'' Penrose limit $A_{11}(u)=A_{22}(u)=0$. In
particular, this happens for radial null geodesics in the Schwarzschild
metric ($p=(1-d)/d$ and $q=2/d$), as already noticed in section 3.1.

These results are summarised in Figure (1a) for $\eta = 1$ and in Figure
(1b) for $\eta = -1$.

\subsection{The Role of the Dominant Energy Condition}

We thus see that while we frequently obtain a singular HPW with
$\omega_a^2 \leq 1/4$ in the Penrose limit, other possibilities do arise.
For timelike singularities, the situation is clear:
\begin{quote}
Penrose Limits of timelike spherically symmetric singularities of power-law
type are singular HPWs with frequency squares bounded from above by $1/4$.
\end{quote}
We will now show that for spacelike singularities a different
behaviour can occur only when the strict Dominant Energy Condition
(DEC) is violated, in particular, that the strongly singular region
(the dark-shaded region in Figure (1a)) is excluded by the requirement
that the DEC be satisfied but not saturated.

We begin by recalling the definition of the {\em Dominant Energy Condition}
on the stress-energy tensor $T^{\mu}_{\;\nu}$ (or Einstein tensor
$G^{\mu}_{\;\nu}$) \cite{HE}: for every timelike vector $v^\mu$,
$T_{\mu\nu}v^{\mu}v^{\nu} \geq 0$, and $T^{\mu}_{\;\nu}v^{\nu}$ 
is a non-spacelike vector. This may be interpreted as saying that for 
any observer the local energy density is non-negative and the energy flux
causal. 

Next we recall that a stress-energy tensor is said to be of type I
\cite{HE} if $T^{\mu}_{\;\nu}$ has one timelike and three (more generally,
$d+1$) spacelike eigenvectors. The corresponding eigenvalues are $-\rho$
($\rho$ the energy density) and the principal pressures $P_{\alpha}$,
$\alpha = 1,\ldots,d+1$.  For a stress-energy tensor of type I, the DEC
is equivalent to
\be
\rho \geq |P_{\alpha}| \;\;.
\label{dec}
\ee
We say that the {\em strict\/} DEC is satisfied if these are strict
inequalities and we will see that the ``extremal'' matter content
(or equation of state) for which at least one of the inequalities is
saturated will play a special role in the following.

The Einstein tensor of the metric (\ref{sim1}) is diagonal (Appendix C),
\bea
G^x_x &=& -\trac{1}{2}d(d-1) x^{-q} - \trac{1}{8}\eta d q
((d-1)q+2p)x^{-(p+2)}    \non
G^y_y &=& -\trac{1}{2}d(d-1) x^{-q} + \trac{1}{8}\eta d q (2p + 4 -
(d+1)q) x^{-(p+2)}    \non
G^i_j &=& -\trac{1}{2}(d-1)(d-2) \d^i_j x^{-q}  + 
\trac{1}{8}\eta (4p- 4q + 4qd - d(d-1)q^2)
\d^i_j x^{-(p+2)}
\eea
and hence clearly of type I. For spacelike singularities, $\eta =+1$, 
we have energy density $\rho=-G^x_x$, radial pressure $P_r=G^y_y$ and
transverse pressures $P_i = G^i_i$, while for $\eta =-1$ the roles of
$G^x_x$ and $G^y_y$ are interchanged. 

Since for $q > p+2$ the first term in $G^x_x$ and $G^y_y$ dominates over
the second term as $x\ra 0$, it is obvious that for $q>p+2$
the relation between $\rho$ and $P_r$ becomes extremal as $x\ra 0$,
\be
G_x^x-G^y_y \ra 0 \;\;\;\;\LRa\;\;\;\;\rho + P_r \ra 0\;\;.
\ee
Put differently, $q \leq p+2$ is a necessary condition for the strict DEC
to hold. Since strongly singular plane waves (the dark-shaded region in
Figure (1a)) arise only for $q>p+2$, we have thus established that
\begin{quote}
Penrose Limits of spacelike spherically symmetric singularities of
power-law type satisfying the strict Dominant Energy Condition are
singular HPWs.
\end{quote}
Since frequency squares exceeding $1/4$ can only occur along the line
$q=p+2$ itself, we can also conclude that
\begin{quote}
the resulting frequency squares $\omega_a^2$ are bounded from above by
1/4 unless one is on the border to an extremal equation of state.
\end{quote}
A more detailed analysis of the DEC (as performed for $d=2$ in \cite{SI}),
shows that the actual region in which the strict DEC is satisfied (taking
into account also the conditions involving the transverse pressures $P_i$),
is more constrained. For spacelike singularities, this is the (infinite)
region bounded by the lines
\be
q=2/d, \,\,\, q = p+2, \,\,\, q= 2(p+1)\;\;,
\ee
displayed as the highlighted region $A$ of Figure (2a) (drawn here for
$d=2$). A look at this figure confirms the results we have obtained
above. 

\begin{figure}
\begin{center}
\includegraphics[width=5 cm]{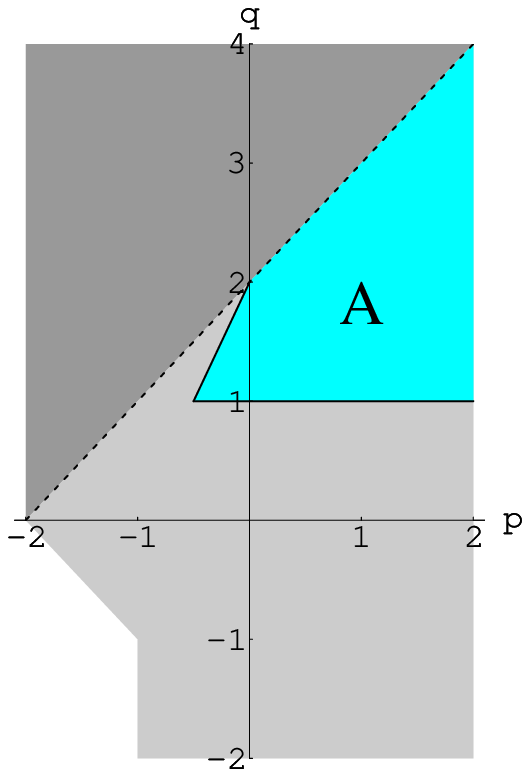}
\hfill
\includegraphics[width=5.3 cm]{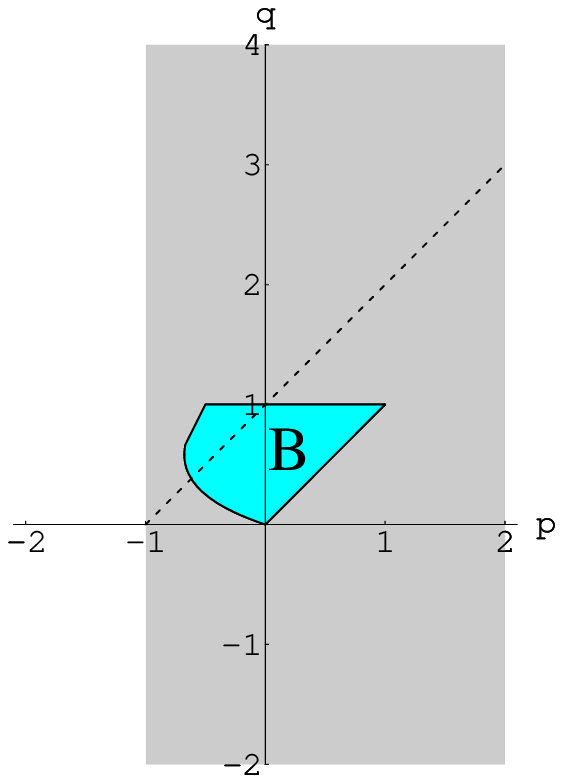}
\end{center}
\caption{The Penrose Limit + DEC Phase Diagram in the $p-q$ plane for (a)
spacelike ($\eta=+1$) and (b) timelike ($\eta=-1$) singularities. In the 
highlighted regions A and B the DEC is satisfied (but not saturated).
(a) the strongly singular (and non-homogeneous) plane waves of the 
dark-shaded region with extremal equation of state are excluded, and 
singular HPWs with $\omega_2^2 \geq 1/4$ arise only along the boundary
$q=p+2$ to the extremal equation of state. (b) the Penrose limits are
singular HPWs with $\omega_2^2 \leq 1/4$, with $\omega_2^2=1/4$ only along
the dashed line $q=p+1$.}
\end{figure}

For timelike singularities, the region where the strong DEC is satisfied
is considerably smaller  - it is a finite subset of the strip bounded by
the lines $q=0$ and $q=2/d$, indicated (for $d=2$) as the highlighted
region $B$ of Figure (2b). While of no consequence for the present
discussion, the fact that in region $B$ the pressures $P_r$ and $P_i$
cannot simultaneously be positive plays an important role in the
discussion of Cosmic Censorship in \cite{SI}.

The fact that the Penrose limits of timelike singularities always behave
as $u^{-2}$, while in the spacelike case strongly singular Penrose
limits can arise (even though only for metrics violating the strict
DEC), might give the impression that timelike (naked) singularities are
in some sense better behaved than spacelike (censored) singularities.
We believe that this should rather be viewed as an indication that
massless particles are inadequate for probing the geometry of timelike
singularities since, for large regions in the $(p,q)$-parameter space,
the angular momentum barrier prevents non-radial null geodesics from
reaching (and hence probing) the singularity. From this point of view,
it is much more significant that for spacelike singularities massless
particle probes with arbitrary angular momentum all detect homogeneous
singular plane waves provided that the strict DEC is satisfied.

We conclude this section with a comment on the case of null singularities
of power-law type ($\eta=0$) which are analysed in \cite{CS}. As
mentioned in section 4.1, in this case some of the leading components
of the Einstein tensor vanish and hence one (somewhat trivially) ends up
with an extremal equation of state. Thus no interesting constraints arise
from imposing the DEC, and using only the leading form (\ref{nullmetric})
of the metric cannot be the basis for a full analysis which is more
subtle and will be left for future work.

\section{Discussion}

We have shown that space-time singularities exhibit a remarkably
universal homogeneous $u^{-2}$-behaviour in the Penrose Limit. We have
established this in complete generality for timelike singularities of
power-law type and have also shown that for spacelike singularities of
power-law type, for which more singular Penrose limits are possible,
this $u^{-2}$-behaviour is implied by demanding the strict DEC.

Perhaps the main implications of this result are for the study of string
theory in singular and/or time-dependent backgrounds. In general,
because of the simplifications brought about by the existence of a
natural light cone gauge \cite{hs}, plane wave (and more general pp-wave)
backgrounds provide an ideal setting for studying such problems.  Now,
as we have seen, the Penrose limits of a large class of singularities
are always at least as singular as $u^{-2}$. Thus ``weakly singular''
plane waves with profile $\sim u^{-\alpha}$, $\alpha < 2$, while perhaps
interesting as toy-models of time-dependent backgrounds in string
theory \cite{sanchez,justin}, do not actually arise as Penrose limits
of standard cosmological or other singularities. Moreover, a strongly
singular behaviour with $\alpha > 2$ can only arise for metrics violating
the strict DEC. This singles out the singular HPWs with profile $\sim
u^{-2}$ as the backgrounds to consider in order to obtain insight into
the properties of string theory near physically reasonable space-time
singularities.

String theory in precisely this class of backgrounds, and in precisely
the frequency range $\omega_a^2 < 1/4$ that we generically find, has already
been studied in some detail and shown to be exactly solvable in
\cite{prt}. The significance of this bound on the frequencies for string
theory (also noted in the earlier studies \cite{sanchez}) lies in the fact
that the properties of Bessel functions allow for an analytic continuation
of string modes through the singularity in this case (due to the standard
shift by 1/4 in the Bessel equation).\footnote{This bound also plays an
important role in a related context in the recent article \cite{gkos}
analysing the propagation of scalar fields in space-time singularities,
where the role of $u$ is played by the usual tortoise-coordinate $r_*$
of Schwarzschild-like metrics.} Other aspects of string theory in this
class of backgrounds still remain (and deserve) to be explored. In
particular, since the Penrose limit can be considered as the origin
of a string expansion around the original background \cite{bfp} (see
\cite{glmssw} for an expansion around the Penrose limit of $AdS_5 \times
S^5$ \cite{bfhp2} in the context of the BMN correspondence \cite{bmn}),
the above observations about the relation of these backgrounds to
interesting space-time singularities provide additional impetus for
understanding string theory in an expansion around such metrics.

The analysis of the present paper can be generalised in various ways.
In particular, even though we have only considered spherically
symmetric metrics, the results we have obtained are certainly not
restricted to spherical symmetry.  A prototypical anisotropic example
is the Kasner metric whose Penrose limit (together with those of other
anisotropic or inhomogeneous cosmological models) was studied by Kunze
\cite{kunze}. In many of these examples one sees that one once again
gets a $u^{-2}$-behaviour either on the nose (for particular Penrose
limits of the Kasner metric) or in the near-singularity limit. It would
be interesting to establish a more general result along these lines, in
particular also in view of the role played by the Kasner metrics (and
Bianchi IX cosmologies) in the BKL discussion of the general solution
of Einstein's equations near spacelike singularities \cite{BKL}.

Going beyond spherical symmetry may also shed light on a peculiar
feature of the spherically symmetric metrics of power-law type studied in
\cite{SI,CS} and here. Namely, we have seen that a rather special role
is played by the ``extremal'' equations of state for which at least one
of the inequalities in the DEC is saturated. As noted in \cite{SI} this
type of equation of state is not ruled out by physical considerations
alone, and arises ``rather too easily'' near the spherically symmetric
singularities of power law type that we have been considering. Indeed,
as we have seen in section 4.4, whenever $q > p+2$ one obtains an
extremal equation of state close to the singularity because then the
``extremal'' contribution to the curvature from the transverse sphere
dominates the longitudinal (radial/time) contribution. Thus this may
well be an artefact of spherical symmetry.

We have also not discussed null singularities in detail. To study these
may not only be of interest in its own right but also because of the
existence of a null counterpart of the BKL solution, a general solution
of the Einstein equations describing a null weak singularity \cite{ori1}.
There is now mounting evidence (see e.g.\ \cite{ori2,bdm}) that these
singularities are realised inside realistic rotating black holes, where
they are expected to form, as a consequence of the Penrose instability,
at the location of what would have been the inner Cauchy horizon of the
stationary Kerr metric. These null singularities bear some resemblance to
the general rotating singular homogeneous plane waves found in \cite{hpw},
and one might wonder if Penrose limits of singularities inside rotating
black holes give rise to these rotating singular homogeneous plane waves
in the same way as rotating smooth (G\"odel) metrics give rise \cite{bmol}
to the rotating smooth homogeneous plane waves of \cite{hpw}. For some
comments on Penrose limits in this context see \cite{ulvi1}.

Since string theory is exactly solvable both in the non-rotating
$u^{-2}$ singular homogeneous plane wave backgrounds \cite{prt} and for
the general rotating smooth homogeneous plane wave metrics \cite{mmga},
one also expects string theory in these general homogeneous singular
backgrounds to be exactly solvable \cite{mmga}, and it will be interesting
to find out if they actually arise from Penrose limits of realistic 
rotating black holes.

\subsection*{Acknowledgements}

This work has been supported by the European Community's Human Potential
Programme under contract HPRN-CT-2000-00131 Quantum Spacetime and by
the Swiss National Science Foundation. MBl is grateful to the High Energy
Group of the Abdus Salam ICTP for hospitality during (what were supposed to
be) the final stages of this work, and MBo thanks CONACyT (Mexico).

\appendix

\section{Hamilton-Jacobi Equations, Geodesic Congruences and Adapted
Coordinates}

Even though the Penrose limit itself only refers to a particular null
geodesic, the original construction of Penrose limits in terms of
adapted coordinates (Penrose coordinates) \cite{penrose,bfp} requires
the embedding of this null geodesic into a twist-free null geodesic
congruence.

Indeed, in Penrose coordinates $(U,V,Y^i)$, $i=1,\dots,
d$, the original space-time metric takes the form
\be
ds^2 = 2dU dV + a(U,V,Y^k) dV^2
+ 2b_i(U,V,Y^k) dV dY^i + g_{ij}(U,V,Y^k)dY^i dY^j\;\;,
\label{pen}
\ee
corresponding to the $(d+2)$ coordinate conditions $g_{UU}=g_{UY^k}=0$, 
$g_{UV}=1$. This exhibits the embedding of the original null geodesic
$V=Y^k=0$ into a congruence of null geodesics 
parametrised by the coordinates $V$ and $Y^k$,
the coordinate $U$ being identified with the affine parameter of
the null geodesics in the congruence. 

A systematic way of constructing Penrose coordinates makes use
of the Hamilton-Jacobi (HJ) function of the null congruence
\cite{patricot,mbictplec}. In this Appendix, we shall explicitly
construct the transformation from a general coordinate system to Penrose
coordinates.

The essence of the HJ method can be summarised by the observation that
the momenta
\be
p_{\mu}=g_{\mu\nu}\frac{dx^{\nu}}{d\lambda}
\ee
associated with the above null congruence $(\dot{U}=1,\dot{V}=\dot{Y}^k=0)$
are
\be
p_{{}_V}=1\;\;,\;\;\;\;\;\;p_{{}_U}=p_{{}_{Y^k}} = 0\;\;,
\ee
so that, in arbitrary coordinates $x^{\mu}$, one has
\be
p_\mu = \del_{\mu} V\;\;.
\la{mmm}
\ee
Thus, since the geodesic congruence is null, 
$g^{\mu\nu} \partial_\mu V \partial_\nu V = 0$,
one can identify
\be
V(x^{\mu}) = S(x^{\mu})\;\;.
\ee
with the solution of the Hamilton-Jacobi equation
\be
g^{\mu\nu} \partial_\mu S \partial_\nu S = 0\;\;.
\label{hjeq}
\ee
corresponding to the null congruence
\be
\dot{x}^{\mu} = g^{\mu\nu}\del_{\nu}S\;\;.
\label{geodeq}
\ee
Conversely (see (\ref{sg})) any solution $S$ of the equations
(\ref{hjeq},\ref{geodeq}) gives rise to a (twist-free) null geodesic
congruence and $V=S$ is the corresponding null adapted coordinate.

It thus only remains to understand how to construct the transverse
coordinates $Y^k$. To that end we will now briefly review some 
facts about solutions to the Hamilton-Jacobi equations (see e.g.\
\cite{goldstein,Benton}).

The general solution to the HJ equation (\ref{hjeq}) can be
rather involved but there usually exists a ``complete'' solution,
complete in the sense that it depends on $d+2$ integration constants
\cite{goldstein} ($d+2$ as in the rest of the paper is the space-time 
dimension).\footnote{It is not always guaranteed that such a
complete solution exists, though in all the cases that we consider
here it does. The most general solution to (\ref{hjeq}) is much more
complicated and can be constructed from a complete solution by looking at
$x$-dependent hypersurfaces in the space of integration constants by the
method of envelopes \cite{Benton}.} For a complete solution to the HJ
equation with integration constants $\alpha_\mu$, the associated geodesic
congruence is $x^\mu= x^\mu(\lambda, \alpha_\mu, x_0^\mu)$, where $x_0^\mu$
are the positions of the geodesics at $\lambda=0$, $x_0^\mu= x^\mu(0,
\alpha_\mu, x_0^\mu)$. The initial value surface parameterised by the
$x_0^\mu$ is a Cauchy surface for the HJ equation and can be represented
algebraically by the equation $F(x^\mu) = 0$.  For a well-posed initial
value problem, we require that the hypersurface $F=0$ has an everywhere
timelike normal vector $(\partial F)^2 <0$.

One of the integration constants $\alpha_\mu$ simply represents a constant
shift of $S$. Furthermore, the HJ equation is homogeneous of degree two,
so if $S$ is a solution, then $\kappa S$, $\kappa=\const\not=0$, is also
a solution.  This scale invariance of the HJ equation is absorbed in the
first order geodesic equations, (\ref{geodeq}) by a scale transformation
of the affine parameter $\lambda$. Therefore, there are only $d$
non-trivial integration constants which we will denote by $\alpha_k$,
$k = 1,\ldots, d$.

Given a particular null geodesic $\gamma$, the integration constants
$\alpha_k$ can be uniquely fixed.  Indeed let $p^0_\mu=g_{\mu\nu}
\dot x^\nu|_{\lambda=0}$ be the momentum of the geodesic $\gamma$
at $\lambda=0$.  The mass-shell condition $g^{\mu\nu}p_\mu p_\nu=0$ is
scale invariant and therefore there are $d$ independent momenta. These
can be used to determine the integration constants of the HJ function $S$
via the equation
\be
p^0_\mu=\partial_\mu S|_{\lambda=0}~.
\ee
Therefore we can use the HJ equation to embed a given null geodesic
into a twist free null geodesic congruence determined by the solution $S$. 

Given a null geodesic $\gamma$, the coordinate transformation from the
original coordinates $x^\mu$ of space-time to the Penrose coordinates
can be defined using the HJ function $S$ and coordinates $x_0^\mu$
of the Cauchy hypersurface, as follows:

We first parameterize the null geodesic congruence as described above
\bea
x^\mu&=&x^\mu(\lambda, x_0^\nu)\\
F(x_0^\mu)&=&0\;\;.
\la{fffeq}
\eea
We have suppressed the  integration constants $\alpha_\mu$ because they are
specified by the momentum of the null geodesic $\gamma$.
Then we set
\bea
U&=& \lambda
\cr
V&=& S(x_0^\mu)\;\;.
\la{uv}
\eea
Note that $S(x^\mu)=S(x_0^\mu)$ because $\dot S=0$.
It remains to determine the coordinates $Y^k$ from these data.
For this observe that the level sets of
$S$ have null normal vector, because of (\ref{hjeq}),  
while the hypersurface $F=0$ has a
timelike  normal vector. Thus we have
\be
g^{\mu\nu}\partial_\mu S \partial_\nu F < 0\;\;,
\ee
and the level sets of $S$  intersect transversally
the hypersurface  $F=0$.
The coordinates $Y^k$ are found by solving the equations
$F(x_0^\mu)=0$ and  $S(x_0^\mu)=V$, i.e.\ $Y^k$ are the coordinates
of the transverse intersection of the $F=0$ hypersurface with the
level sets of the HJ function $S$.
Using this, we can rewrite the first equation in (\ref{fffeq}) as
\be
x^\mu=x^\mu(U, x_0^\nu(V,Y^k))=x^\mu(U, V,Y^k)\;\;.
\ee
This is the transformation which relates a coordinate
system on a space-time to the Penrose coordinates.

Note that for a generic space-time, there is no natural choice for the
hypersurface $F=0$, i.e.\ for the function $F$.  Instead $F$ should be
thought of as a gauge fixing condition which is chosen at our convenience.
The Penrose limit metric does not depend on the choice of $F$, different
choices simply corresponding to different ways of labelling the geodesics
of the congruence on which the adapted coordinates are based.

For the sake of completeness we will now show explicitly that in these
coordinates the metric indeed takes the form (\ref{pen}) 
\cite{patricot,mbictplec}. First of all, we clearly have
\be
g_{UU}=g_{\mu\nu} {\partial x^\mu\over\partial \lambda}
{\partial x^\nu\over \partial \lambda}=0
\ee
because the geodesics $x^\mu(\lambda, x_0^\nu)$ are null. Moreover,
\be
g_{UV}= g_{\mu\nu} {\partial x^\mu\over\partial \lambda} 
{\partial x^\nu\over\partial V}=
g_{\mu\nu} g^{\mu\rho}\partial_\rho S {\partial x^\mu\over\partial V}
= {\partial x^\mu\over\partial V} \partial_\mu S=
{\partial V\over\partial V}=1\;\;,
\ee
and
\be
g_{U i}= 
g_{\mu\nu} {\partial x^\mu\over\partial \lambda} 
{\partial x^\nu\over\partial Y^i}=
g_{\mu\nu}g^{\mu\rho}{\partial S\over \partial x^\rho} 
{\partial x^\nu\over\partial Y^i}=
{\partial V\over \partial Y^i}=0\;\;.
\ee

\section{Generalisations: Brane Metrics, Isotropic Coordinates, Null
Singularities}

It is straightforward to generalise the analysis of section 2.3 
to include longitudinal worldvolume directions,
\be
f(r)(-dt^2) \ra f(r) (-dt^2 + d\vec{y}^2)\;\;.
\ee
A parallel frame in the brane worldvolume directions 
is $E_i = f^{-1/2} \del_{y^i}$, and
\be
B_{ij} = \d_{ij} \del_u \log f(r(u))^{1/2}
\ee
which in turn leads to
\be
A_{ij} = \d_{ij}
f(r(u))^{-1/2} \del_u^2 \log f(r(u))^{1/2}\;\;.
\ee
The remaining of the components of $A$ are as in section 2.3.

Likewise, for isotropic coordinates,
\be
ds^2  = -f(r) dt^2 + h(r) (dr^2 + r^2 d\Omega_d^2)\;\;,
\ee
a straightforward calculation reveals that
\be
\tr B =
\partial_u\log \Biggl(\dot r r^d f^{\frac{1}{2}} 
h^{\frac{d+1}{2}} \sin^{d-1}(\theta)\Biggr)
\ee
and
\be
B_{22}=\partial_u \log\bigl( h^{\frac{1}{2}} r \sin(\theta)\bigr)~.
\ee
These lead to
\bea
B_{11}(u) &=& \del_u \log \bigl(r \dot{r} h f^{1/2}\bigr)
\non
B_{\hat{a}\hat{b}}(u) &=& \delta_{\hat{a}\hat{b}}
\del_u \log (r h^{1/2}\sin\theta)
\eea
with the corresponding second-derivative expressions for $A_{ab}(u)$.
Again it is easy to include longitudinal directions.

Finally, we consider spherically symmetric null metrics
of the form
\be
ds^2 =2 g(x) dx dy + f(x) d\Omega_2^2~.
\ee
The geodesic equations are
\be
\dot x=P g^{-1}\;\;,\;\;\;\; 
\dot y= -\frac{L^2}{P} f^{-1}\;\;,\;\;\;\; \dot \theta=f^{-1} L\;\;,
\label{gogo}
\ee
where $P$ and $L$ are constants of motion. In this case, one finds
\be
\tr B =
\partial_u \log\bigl( \dot x g f \sin(\theta)\bigr)
\ee
and
\be
B_{22}=\partial_u\log\bigl(f^{\frac{1}{2}} \sin(\theta)\bigr)~.
\ee
Therefore, we have
\bea
K_1(u) &=& P f(u)^{1/2}\non
K_2(u) &=& f(u)^{1/2}\sin\theta(u)\;\;.
\eea
In particular, in terms of $r(x) = f(x)^{1/2}$,~ $A_{22}$ once again
takes the standard form (\ref{a22})
\be
A_{22}(u)=\frac{\ddot{r}(u)}{r(u)}-\frac{L^2}{r(u)^4}\;\;.
\ee

\newpage

\section{Curvature of Szekeres-Iyer Metrics}

For reference purposes we give here the non-vanishing components of the
Ricci and Einstein tensors of the metric (\ref{sim1}),
\be
ds^2 = \eta x^p dy^2 - \eta x^p dx^2 + x^q d\Omega_d^2 
\ee
(for $d=2$, these
results can be inferred from \cite{SI}). Indices $i,j$ refer to the 
metric $\hat{g}_{ij}$ of the transverse sphere (or some other transverse
space), with $\hat{R}_{ij}$ and $\hat{R}$ the corresponding Ricci tensor 
and Ricci scalar. 

\subss{Ricci Tensor}
\bea
R_{xx} &=& \trac{1}{4}(2p + 2qd + pqd - q^2 d) x^{-2}\non
R_{yy} &=& \trac{1}{4}p (qd - 2) x^{-2}\non
R_{ij} &=& \hat{R}_{ij} + \trac{1}{4}\eta q (q d - 2)\hat{g}_{ij}
x^{q-p-2}\non
 &=& (d-1)\hat{g}_{ij} + \trac{1}{4}\eta q (q d - 2)\hat{g}_{ij} x^{q-p-2}
\eea

\subss{Ricci Scalar}
\bea
R &=& \hat{R}x^{-q} - \trac{1}{4}\eta (4p + 4qd - d(d+1) q^2) x^{-(p+2)}\non
&=& d(d-1) x^{-q} - \trac{1}{4}\eta (4p + 4qd - d(d+1) q^2) x^{-(p+2)}
\eea

\subss{Einstein Tensor}
\bea
G^x_x &=& -\trac{1}{2}\hat{R} x^{-q} - \trac{1}{8}\eta d q
((d-1)q+2p)x^{-(p+2)}    \non
&=& -\trac{1}{2}d(d-1) x^{-q} - \trac{1}{8}\eta d q
((d-1)q+2p)x^{-(p+2)}    \non
G^y_y &=&     -\trac{1}{2}\hat{R} x^{-q} + \trac{1}{8}\eta d q (2p + 4 -
(d+1)q) x^{-(p+2)}    \non
&=&     -\trac{1}{2}d(d-1) x^{-q} + \trac{1}{8}\eta d q (2p + 4 -
(d+1)q) x^{-(p+2)}    \non
G^i_j &=& \hat{G}^i_j x^{-q} + \trac{1}{8}\eta (4p - 4q + 4qd - d(d-1)q^2)
\d^i_j x^{-(p+2)}\non
&=&   -\trac{1}{2}(d-1)(d-2) \d^i_j x^{-q}  + 
\trac{1}{8}\eta (4p- 4q + 4qd - d(d-1)q^2)
\d^i_j x^{-(p+2)}
\eea

\newpage

\rnc{\Large}{\normalsize}

\end{document}